# Manufacturing Pathway and Experimental Demonstration for Nanoscale Fine-Grained 3-D Integrated Circuit Fabric

Mostafizur Rahman, Jiajun Shi, Mingyu Li, Santosh Khasanvis, Csaba Andras Moritz*
Electrical and Computer Engineering, University of Massachusetts Amherst, Amherst, USA
andras@ecs.umass.edu*

*Summary*—At Sub-20nm technologies CMOS scaling faces severe challenges primarily due to fundamental device scaling limitations, interconnection overhead and complex manufacturing. Migration to 3-D has been long sought as a possible pathway to continue scaling; however, CMOS's intrinsic requirements are not compatible for fine-grained 3-D integration. In [1], we proposed a truly fine-grained 3-D integrated circuit fabric called Skybridge that solves nanoscale challenges and achieves orders of magnitude benefits over CMOS. In Skybridge, device, circuit, connectivity, thermal management and manufacturing issues are addressed in an integrated 3-D compatible manner. At the core of Skybridge's assembly are uniform vertical nanowires, which are functionalized with architected features for fabric integration. All active components are created primarily using sequential material deposition steps on these nanowires. Lithography and doping are performed prior to any functionalization and their precision requirements are significantly reduced. This paper introduces Skybridge's manufacturing pathway that is developed based on extensive process, device simulations and experimental metrology, and uses established processes. Experimental demonstrations of key process steps are also shown.

*Extended Abstract*—As CMOS is reaching its fundamental limits at nanoscale, fine-grained 3-D CMOS to continue scaling has been extremely difficult to achieve due to inherent doping and customization requirements of CMOS circuits. Partial 3-D attempts with die-die and layer-layer stacking only provide incremental density benefits while retain CMOS scaling challenges, and have their own limitations[2][3]. Skybridge is a new truly fine-grained 3-D integrated fabric that provides an integrated solution for all nanoscale technology aspects, and achieves orders of magnitude benefits [1]. In Skybridge's 3-D integration, uniform vertical nanowires play a key role. Vertical Gate-All-Around (V-GAA) Junctionless transistors that do not require doping variations within the device, are active devices and are realized through material depositions on nanowires. V-GAA Junctionless transistors are interconnected through nanowire connecting Bridges and nanowire surrounding routing structures in a 3-D circuit style that uses only single-type, uniformly sized transistors for logic and memory circuits. Fabric's intrinsic heat management is through heat extraction features and heat dissipating nanowires. Since, formation of all active fabric components is by functionalizing nanowires through material depositions, lithographic precision is required only for nanowire patterning. Moreover, Skybridge's unique circuit design approach with uniform Junctionless transistors ensures doping is required only once during initial wafer preparation, reducing manufacturing constraints further.

Brief manufacturing pathway with key steps for fabric assembly are depicted in Fig 1B, and a cross-section of 3-D circuit (3-input 3-D NAND gate) with underlying materials is shown in Fig. 1A. The manufacturing pathway is derived based on process, device simulations and our experimental demonstration of Junctionless transistor in 2-D [4], and uses existing foundry processes. The process flow starts with substrate doping (Fig. 1B(i)), and is followed by nanowire patterning (Fig. 1B(ii)). Subsequent steps involve controlled material depositions on these nanowires. Contact regions are formed through anisotropic metal (Ti) deposition at the bottom with sacrificial layer (Polyimide) on top (Fig. 1B(iii)). This step is followed by metal deposition (W) for interconnect layer (Fig. 1B(iv)), planarization with dielectric (SU-8) etch back, and spacer ($Si_3N_4$) deposition (Fig. 1B(v)). Low-$k$ dielectric material (C-doped $SiO_2$) is deposited next, and is followed by planarization (Fig. 1B(vi)). For transistor's Gate formation, $HfO_2$ is deposited next using ALD (Fig. 1B(vii)), followed by Gate material (TiN) deposition (Fig. 1B(viii)). After TiN deposition, $HfO_2$ is etched back; Gate formation is completed by contact material (W) depositions (Fig. 1B(ix)). Using 3-D TCAD process and device simulators, the physical process flow was emulated for device creation and carrier transport was simulated. Figs. 1C and 1D shows device cross-section with exact material dimensions used in simulations, and simulated results, respectively. Fig. 1D shows excellent device characteristics with $10^5$ $I_{on}/I_{off}$, and 0.3V $V_{th}$.

Following the manufacturing pathway, we have experimentally demonstrated key steps necessary for fabric assembly. These include: high-aspect ratio nanowire array patterning, photoresist planarization, anisotropic material deposition, surface planarization, and multi-layer selective deposition. Key results are shown through Figs. 1E-1H. Arrays of nanowires that are 1100nm tall and 197nm wide were patterned using E-beam, and RIE (Fig. 1E). Using alignment techniques, selective nanowire regions were exposed, and metal (Ti) was anisotropically deposited for Contact formation (Fig. 1F). This step was followed by surface planarization using dielectric (SU-8) over-fill and etch-back (Fig. 1G). Finally, selective regions were exposed again for multi-layer deposition demonstrations (Fig. 1H). The full paper will include all manufacturing details.

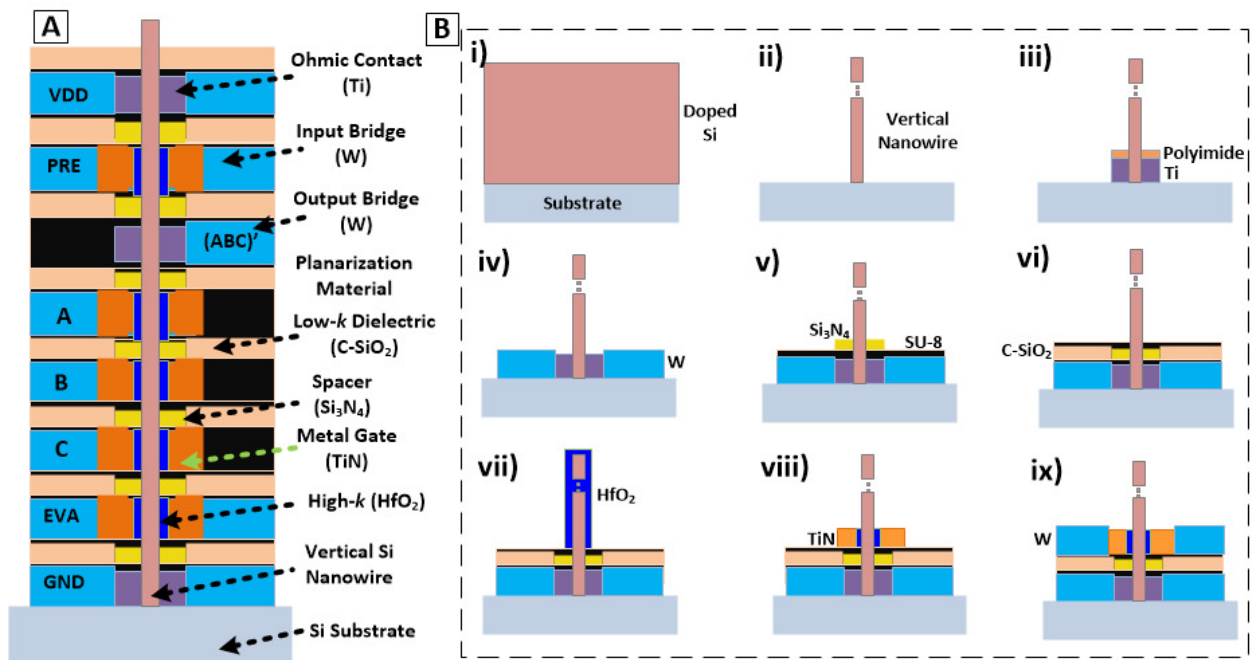

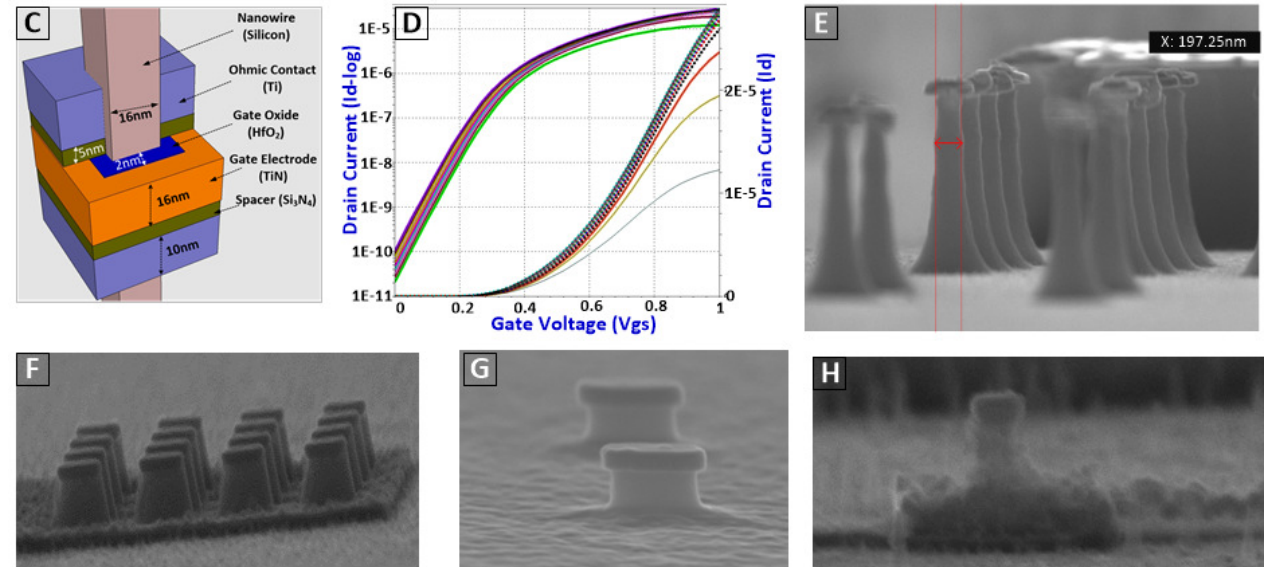

Fig. 1. A) Cross-section of a 3-D circuit in Skybridge fabric. B) Key steps for fabric assembly; i) substrate doping, ii) vertical nanowire patterning, iii) anisotropic material (Ti) deposition at the bottom with sacrificial layer, iv) formation of Bridges (W), v) surface planarization (SU-8) and spacer ($Si_3N_4$) deposition, vi) low-$k$ inter-layer dielectric (C-$SiO_2$) deposition followed by planarization, vii) ALD deposition of $HfO_2$, viii) Gate material (TiN) deposition, ix) Gate contact (W) formation. C) Cross-section of V-GAA Junctionless transistor. D) 3D TCAD simulated device characteristics. E) Fabricated nanowire arrays; each nanowire is 1100nm tall, and 197nm in width. F) Demonstration of selective material deposition anisotropically. G) Planarization with SU-8. H) Multi-layer selective depositions.